\documentclass[aps,amsmath,amsfonts,superscriptaddress,showpacs,twocolumn,prb,10pt]{revtex4-1}
\usepackage{graphicx}
\usepackage{epsfig}
\usepackage{bm}
\usepackage{dcolumn}
\usepackage{amsmath}
\usepackage{amssymb}
\usepackage{color}

\newcommand{\ud}{\,\mathrm{d}}
\newcommand{\im}{\textrm{i}}

\newcommand{\ver}{{\vec{r}}}

\DeclareMathOperator{\sgn}{sgn}
\DeclareMathOperator{\tr}{tr}
\DeclareMathOperator{\e}{e}
\DeclareMathOperator{\F}{\mathcal{F}}
\DeclareMathOperator{\T}{\mathcal{T}}

\begin{document}

\title{Scanning tunneling microscopy of superconducting topological surface states in $\textrm{Bi}_2\textrm{Se}_3$}

\author{Ian M. Dayton} 
\email{ian.m.dayton@gmail.com}
\affiliation{Department of Physics and Astronomy, Michigan State University, East Lansing, Michigan 48824, USA}
\author{Nicholas Sedlmayr}
\affiliation{Department of Physics and Astronomy, Michigan State University, East Lansing, Michigan 48824, USA}
\author{Victor Ramirez} 
\affiliation{Department of Physics and Astronomy, Michigan State University, East Lansing, Michigan 48824, USA} 
\author{Thomas Chasapis} 
\affiliation{Department of Chemistry, Northwestern University, Evanston, Illinois 60208, USA}
\author{Reza Loloee}  
\affiliation{Department of Physics and Astronomy, Michigan State University, East Lansing, Michigan 48824, USA}
\author{Mercouri Kanatzidis}  
\affiliation{Department of Chemistry, Northwestern University, Evanston, Illinois 60208, USA} 
\author{Alex  Levchenko}
\affiliation{Department of Physics, University of Wisconsin-Madison, Madison, Wisconsin 53706, USA}
\author{Stuart Tessmer} 
\email{tessmer@pa.msu.edu} 
\affiliation{Department of Physics and Astronomy, Michigan State University, East Lansing, Michigan 48824, USA}

\begin{abstract}
In this paper we present scanning tunneling microscopy of a large $\textrm{Bi}_2\textrm{Se}_3$ crystal with superconducting PbBi islands deposited on the surface. Local density of states measurements are consistent with induced superconductivity in the topological surface state with a coherence length of order 540 nm. At energies above the gap the density of states exhibits oscillations due to scattering caused by a nonuniform order parameter. Strikingly, the spectra taken on islands also display similar oscillations along with traces of the Dirac cone, suggesting an inverse topological proximity effect.
\end{abstract}

\pacs{74.45.+c, 74.55.+v, 03.65.Vf} 

\date{\today}

\maketitle

\textbf{\textit{Introduction}}. Three-dimensional topological insulators (TI) \cite{Hasan-Kane}, such as $\textrm{Bi}_2\textrm{Se}_3$, Bi$_2$Te$_3$ or Bi$_{1-x}$Sb$_x$, were once only known for having great thermoelectric properties. Their most notable physical characteristic, to harbor gapless surface states with striking spin textures protected by topology, was discovered only recently \cite{Hsieh-BiSb,Xia-BiSe,Zhang-BiSe-BiTe}. It is understood nowadays that these topological surface states (TSS) stem from a combination of strong spin-orbit coupling and band inversion in these materials. It has been quickly realized that studying the interplay of such symmetry-protected states and symmetry-broken phases, including for example magnetism or superconductivity, may lead to a plethora of new effects and provide new platforms for potential technological advances. The most natural approach to achieve this goal is to study a set of phenomena associated with the proximity effect. 

It has been predicted that when a topological insulator is brought into contact with a conventional $s$-wave superconductor (SC) the proximity effect induces superconducting correlations into the TSS that have unconventional $p$-wave symmetry \cite{Fu-Kane}. This original result triggered a flood of further theoretical works, some representative examples include Refs.~\onlinecite{Linder,Potter,Lababidi,Ioselevich,Tkachov,Reeg}, and a multitude of experimental efforts. The latter span from angle-resolved photoemission spectroscopy, scanning tunneling microscopy, point contact, and differential conductance measurements \cite{Sasaki,Wang-BiSe-NbSe-STM,Wang-BiSe-BiSrCaCuO-ARPES,Koren-BiSe-NbN,Xu-BiTe-STM}, through to observations of supercurrents and unusual Josephson Fraunhofer patterns \cite{Sacepe,Veldhorst,Qu,Bestwick,Mason,Kurter-PRB,Kurter-NC,Sochnikov,Stehno}; experiments on phase coherent transport including multiple Andreev reflections, Fabry-Perot interferometry and Aharonov-Bohm oscillations \cite{Zhang-PRB11,Zareapour,Finck-PRX,Finck-arXiv}.
       
\begin{figure}
\includegraphics[width=\linewidth]{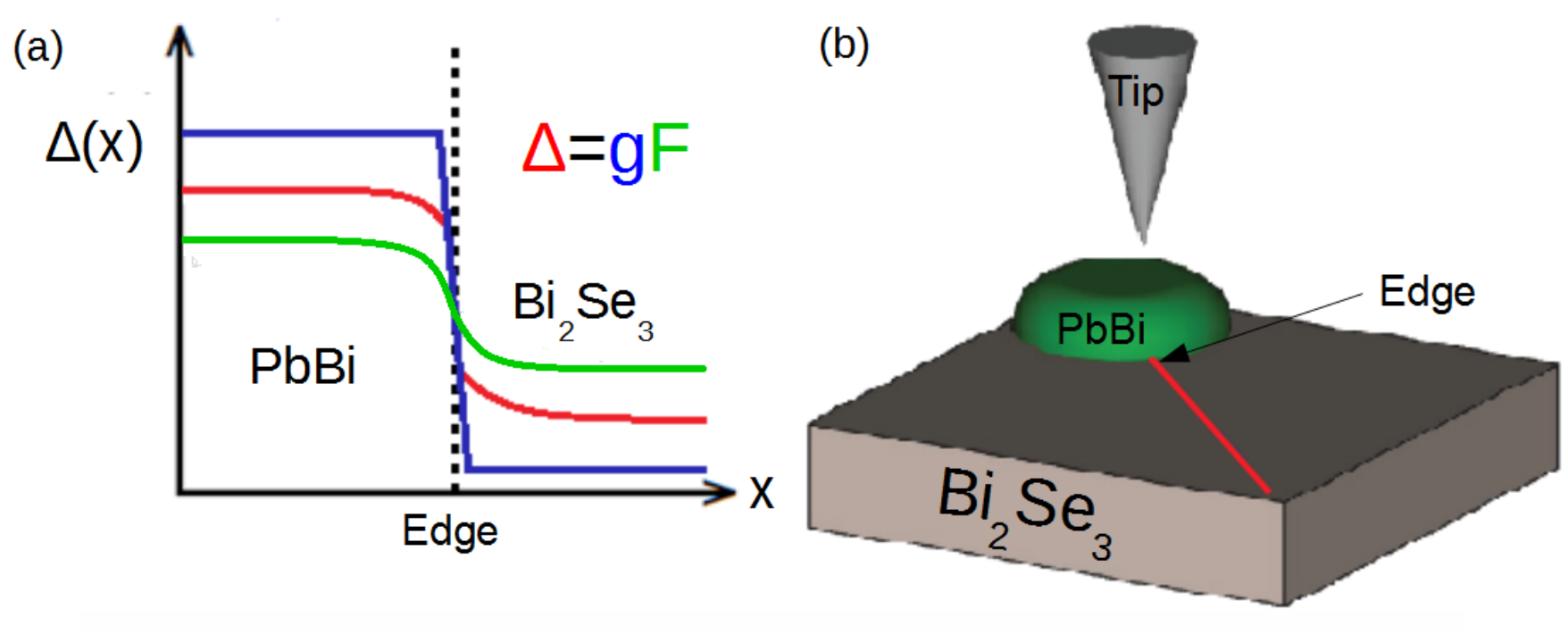}
\caption{(Color online) (a) A schematic of the standard proximity effect picture showing the induced superconducting energy gap\cite{degennes}, $\Delta$, at an interface as a function of position $x$, where $F$ is the superconducting condensate amplitude. The pairing interaction constant, $g$, is generally taken to have the form of a clear step function, but at small scales, the step actually has a finite slope due to electronic screening \cite{DDG}. (b) A schematic illustration for the geometry of the STM probe scanning over the surface of TI $\textrm{Bi}_2\textrm{Se}_3$ in proximity to a PbBi island. In this experiment, we are interested in measuring the superconducting coherence length in the plane parallel to the TI surface, not in the normal direction to the surface as done in many other experiments.}
\label{fig:Figure1}
\end{figure}
       
On the technical side, proximity effects can be realized by growing thin layers of SC-TI heterostructures, an approach employed by most of the existing experiments. Alternatively, one could study induced superconductivity in the surface state from the TI's bulk which becomes superconducting by Cu-intercalation into Bi$_2$Se$_3$ \cite{Wray,Hor}. While Cu$_x$Bi$_2$Se$_3$ retains the Dirac surface state, its superconducting volume fraction is relatively low which causes obvious challenges. Here we explore a complimentary but different route inducing superconductivity into TSS locally by depositing a matrix of superconducting PbBi islands on the surface of Bi$_2$Se$_3$. This approach was theorized earlier as a path to creating superconducting graphene \cite{Tikhonov}. It was employed experimentally to create a tunable realization of two-dimensional superconductivity in mesoscopic superconductor-normal-superconductor arrays \cite{Mason-SNS}. 

Through the use of cryogenic scanning tunneling microscopy (STM), an experimental technique that is ideal for probing the superconducting proximity effect in inhomogeneous samples, we are able to measure the induced local electronic density of states in TI surface states. A schematic illustration of the experiment is shown in Fig.~\ref{fig:Figure1}(b), where an STM probe is placed near the interface of a superconducting island and a TI, and spectra are taken while gradually moving the probe away from the island along the TI surface. 

We report in this letter the observation of two main experimental results. First, we show that STM spectra reveal a clear superconducting gap induced into the TSS. From the spatially resolved probes and fits of the gap function we estimate the superconducting coherence length to be of the order of $\lesssim 540 \textrm{ nm}$ along the direction parallel to the quintuple layers. In addition, at energies above the gap we observe oscillatory behavior of the density of states that resembles the Tomasch interference effect \cite{Wolfram}. Second, while all the existing efforts were concentrated on revealing signatures of superconductivity induced into the TSS, much less attention was paid to the corresponding reverse effect of the TSS on an adjacent superconductor. We address this intriguing question by taking careful STM density of states spectra on superconducting islands and uncovering traces of the Dirac cone that seemingly leaks from the TSS. This observation manifestly provides evidence for the possibility of an inverse topological proximity effect.      

\begin{figure}
\includegraphics[width=\linewidth]{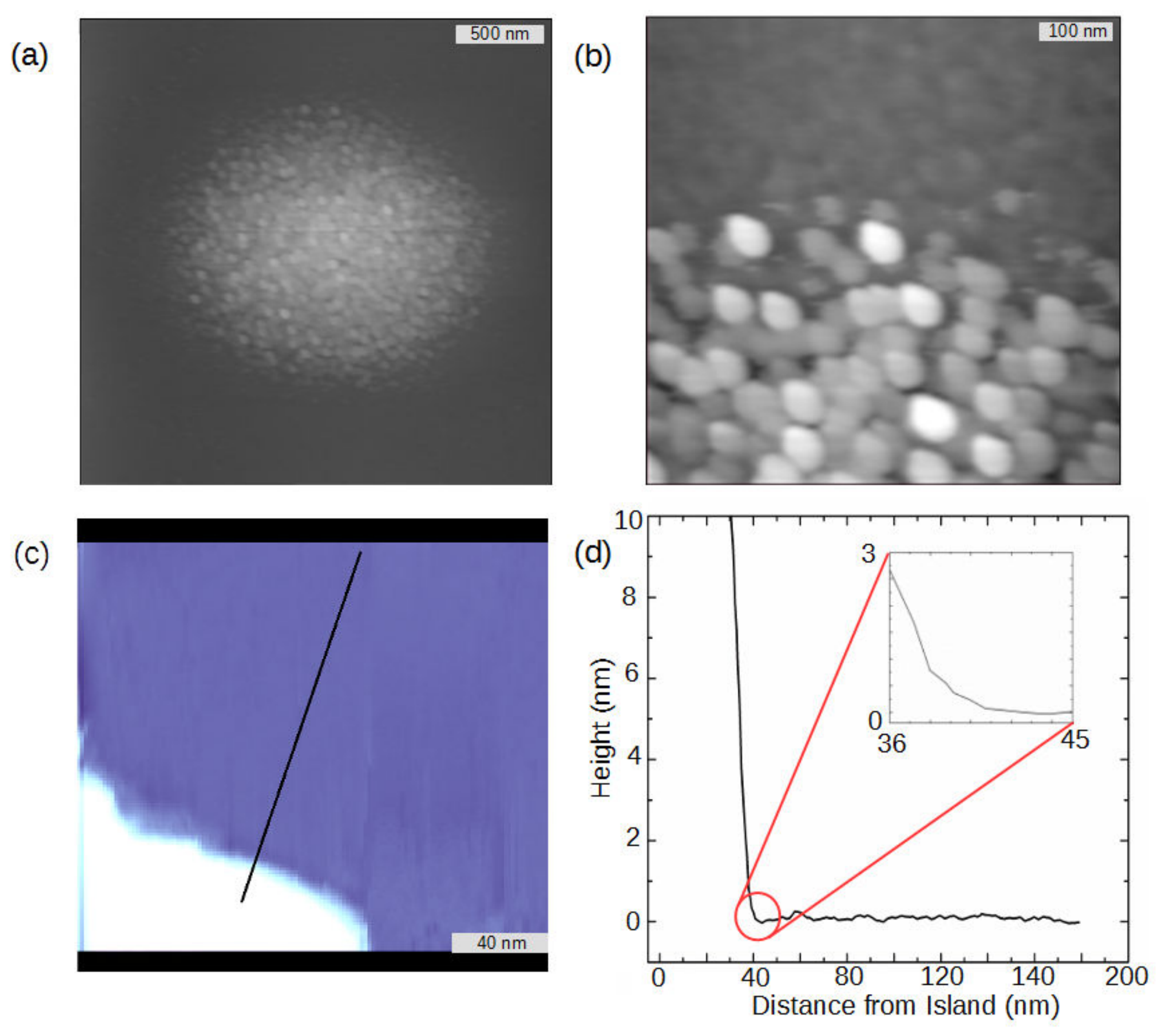}
\caption{(Color online) (a) and (b) Atomic force microscopy (AFM) topographs of a thermally deposited PbBi dot.  We can clearly see a very grainy appearance to the overall dot, but upon closer inspection it is clear that the dot is comprised of many small superconducting droplets. (c) STM topograph of a PbBi droplet with its respective height profile trace shown in (d). The radius of curvature at the base of the droplet is an artifact due to the radius of the STM tip.  The actual interface between the droplet and the TI is very sharp, which is ideal for our measurement.}
\label{fig:Figure2}
\end{figure}

\textbf{\textit{Measurements and results}}. The TI used in this experiment is Bismuth doped $\textrm{Bi}_{2.04}\textrm{Se}_{2.96}$, grown by slowly cooling a stoichiometric mixture of Bi and Se from a temperature of $850 \textrm{ }^{\circ} \textrm{C}$.  Five atomic planes with atomic order Se1-Bi1-Se2-Bi1-Se1 form a quintuple layer (QL); the QLs are weakly bound to each other, making it possible to readily expose a pristine surface for study. The exposed QL supports the existence of the TSS, which features a single Dirac cone.  While $\textrm{Bi}_{2}\textrm{Se}_{3}$ is typically $n$ type, the bulk doping of Bi tends to shift the Fermi level back to the center of the band gap \cite{Megan}.  $\textrm{Pb}_{0.3}\textrm{Bi}_{0.7}$ is used as the superconductor due to its favorable whetting properties on $\textrm{Bi}_{2}\textrm{Se}_{3}$, its large gap width (3.65 meV), and its high transition temperature of 8.2 K \cite{Adler}.  

The TI is first cleaved in a nitrogen environment then placed into a thermal evaporator.  Here, we deposit 10 nm of $\textrm{Pb}_{0.3}\textrm{Bi}_{0.7}$ onto the surface of the TI by evaporation through a (transmission electron microscopy) TEM mask.  This results in large array of superconducting islands with a diameter of 1.2 $\mu\textrm{m}$.  From there, the sample is moved to our cryogenic Besocke design STM system for measurement.  All of these steps are taken in a vacuum, nitrogen, or helium environment, so the sample is never exposed to air.  All STM topographs are taken with a bias voltage of 5 V and tunneling current of 500 pA, and all spectra are taken over a range of -45 meV to 45 meV and measured via a lock-in amplifier.

When doing a surface probe measurement, it is important to fully characterize the surface being measured.  In this case, we are interested in the quality of deposition of the PbBi islands as well as how clean the interface is between the island edge and the TI surface.  In Fig.~\ref{fig:Figure2}, we show various atomic force microscopy (AFM) and STM topographs demonstrating the structure of these islands after thermal evaporation, we want to stress that these samples were also not exposed to the air.  The PbBi dots appear to be comprised of many  20-100 nm radius droplets grown on top of and around one another.  This can clearly be seen in Fig.~\ref{fig:Figure2}(a) and (b).  Fig.~\ref{fig:Figure2}(c) is an STM topograph showing the edge of one of such droplet formations, along with the respective height profile in Fig.~\ref{fig:Figure2}(d). Here we can see that the interface between the PbBi and the TI surface is very abrupt, giving evidence of minimal leakage of the PbBi onto the TI surface. 
   
\begin{figure}
\centering
\includegraphics[width=\linewidth]{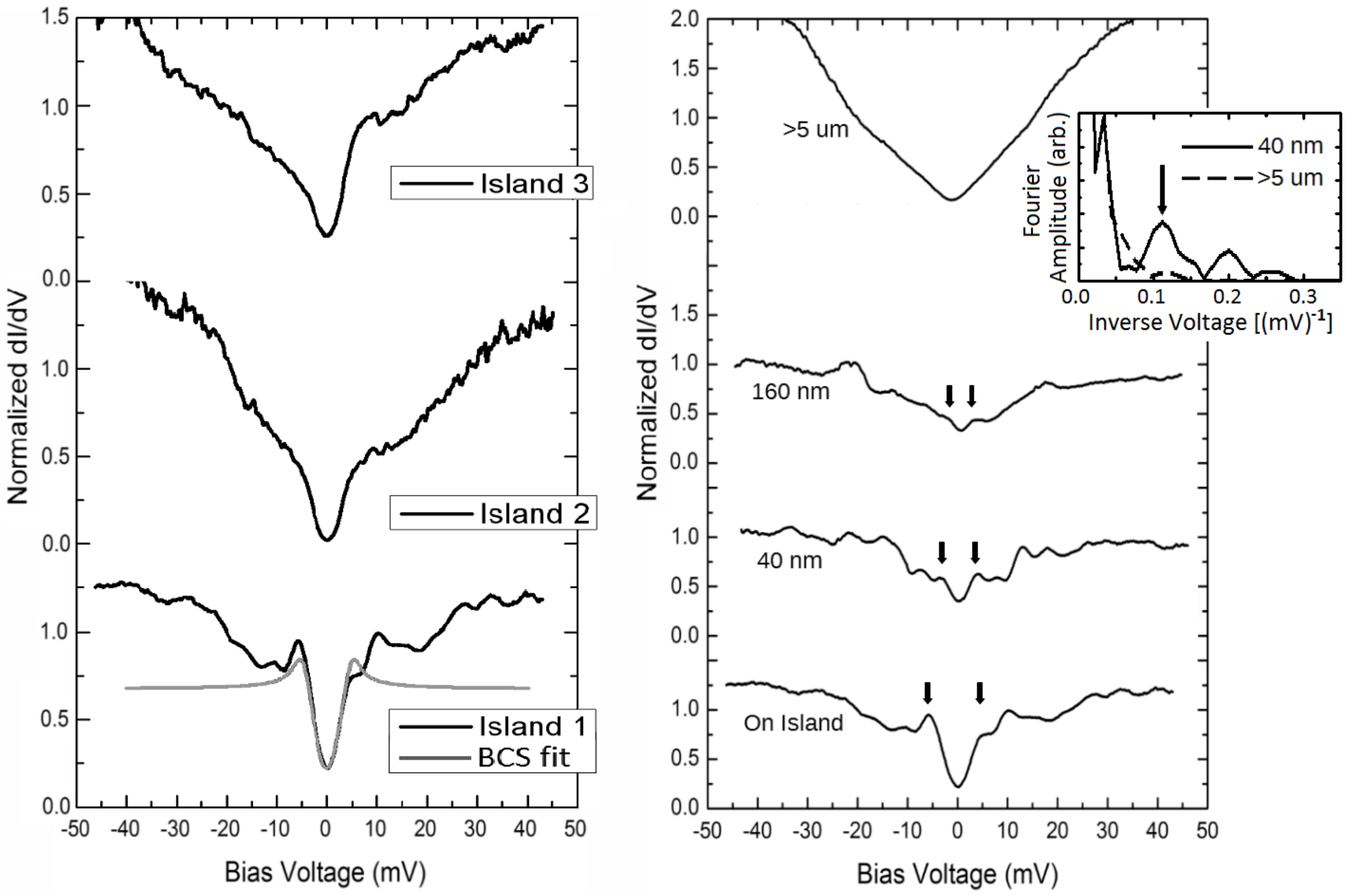}
\caption{The right panel represents $dI/dV$ curves measured at 4.2 K taken at various distances from a PbBi island. The LDoS displays clear signature of the induced superconducting gap. Another notable feature of the presented data is visible oscillations, the frequency of oscillations is visible in the Fourier transforms shown in the inset. The left panel represents $dI/dV$ curves measured on different islands at nominally the same conditions. Above the gap one can see not only oscillations but also traces of the Dirac cone at higher energies. All data are normalized so that $\ud I/\ud V|_{-20\textrm{mV}}=1$. The asymmetry which can be seen in the top left two panels is due to an offset between the Fermi energy and the minimum of the Dirac cone caused by the states leaking onto the superconducting island.}
\label{fig:Figure3}
\end{figure}

We present the local density of states (LDoS) measurements taken via cryogenic STM at a temperature of 4.2K in Fig.~\ref{fig:Figure3}. The right panel shows a series of differential conductance plots reflecting LDoS of TSS taken at various distances away from the superconducting island. At a distance of approximately 40 nm the superconducting induced gap is roughly 20\% smaller than the corresponding gap on an island, while at distance of order 200 nm the gap falls to almost half of its value. At a distance of $>5$ $\mu \textrm{m}$, the Dirac cone is the only dominating feature in the LDoS since the local region of the TI is no longer within the range of the superconducting proximity effect.On the data set taken at 40 nm away from the island one sees oscillatory features occurring with the period of 5-10 mV. As seen in the inset of Fig. 3, the magnitude of the Fourier transform shows a clear resonance at 0.11 $mV^{-1}$, corresponding to a periodicity of 9 mV; a near harmonic is evident at 0.20 $mV^{-1}$. Interestingly, similar features are present on the LDoS plots taken on the different islands, which are shown on the left panel of Fig.~\ref{fig:Figure3}. Furthermore, in addition to oscillations at energies above the gap, traces of the Dirac cone are also revealed. We attribute this dependence to the inverse topological proximity effect where LDoS properties of topological surface states penetrate into the superconducting island. In what follows we illustrate this behavior by a simple theoretical model. 

\textbf{\textit{Discussion and analysis}}. It has been emphasized early on that crystals of Bi$_2$Se$_3$ possess topological surface states, as well as intra-gap trivial states in the bulk originating from unintentional doping, and importantly also at the surface due to the band bending effect \cite{Bianchi}. This coexistence of topological and trivial surface states leads to certain complications in the context of the proximity effect. In particular, phase coherent transport measurements \cite{Bestwick,Finck-PRX,Finck-arXiv} suggest that the superconducting proximity effect of trivial states is in the diffusion dominated transport regime, whereas topological states display transport peculiarities that are specific to the ballistic domain of transport.   

\begin{figure}
\centering
\includegraphics[width=\linewidth]{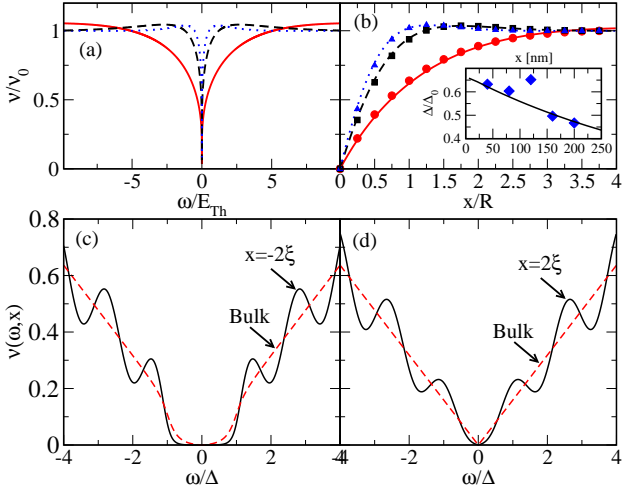}
\caption{(Color online) (a,b) The local density of states in the diffusive limit as a function of energy (a) at different distances from the boundary of $0.5R, 1R, 1.5R$, and as a function of position (b) at energies $0.1E_{\rm Th},0.8E_{\rm Th},1.5E_{\rm Th}$, with lines representing the linearized analytical result, and symbols the numerical result of full nonlinear Eq.~\eqref{Usadel}. The inset show a comparison between the gap found experimentally (blue diamonds) and the analytical result (solid line). (c,d) The local density of states as a function of energy for the topological insulator surface states. A phenomenological damping has been included, and the bulk density of states are included as a comparison.}
\label{fig:Figure4}
\end{figure}

In order to gain some theoretical insight into the system we have modeled our experimental set-up in two limits. As a first approach we have considered the diffusive limit that should be relevant to a part of the proximity effect governed by the trivial surface states. In complete analogy to previously studied mesoscopic superconductor-normal proximity junctions \cite{SN-STM,AL} we solved the standard Usadel equation for a circular geometry describing a superconducting island of radius $R$ surrounded by an infinite normal system. Within this formalism the proximity effect is described by the semiclassical Green's function $G(x,\omega)=\cos[\theta(x,\omega)]$ of position and energy that in the normal region obeys the nonlinear equation 
\begin{equation}\label{Usadel}
\frac{\partial^2\theta}{\partial x^2}+\frac{1}{x}\frac{\partial\theta}{\partial x}+\frac{i\omega}{E_{\mathrm{Th}}}\sin\theta=0
\end{equation}
where $x=r/R$ is the dimensionless position coordinate and $E_{\mathrm{Th}}=v_Fl/2R^2$ is the Thouless energy defined by the Fermi momentum $v_F$ and mean free path $l$. The LDoS is obtained from the real part of the Green's function 
\begin{equation}\label{LDoS-Usadel}
\nu(x,\omega)=\nu_0\Re\cos[\theta(x,\omega)],
\end{equation}
where $\nu_0$ is the normal density of states without the proximity effect. 
In a linearized regime, applicable at distances away from the boundary where the proximity effect is weak, an analytical result for Eq.~\eqref{Usadel} is possible,
\begin{equation}\label{theta}
\theta(x,\omega)=\theta_0(\omega)\frac{K_0(x\sqrt{i\omega/E_{\mathrm{Th}}})}{K_0(\sqrt{i\omega/E_{\mathrm{Th}}})},
\end{equation} 
elsewhere the problem must be solved numerically. In Eq.~\eqref{theta} $K_0(z)$ is the modified Bessel function and $\theta_0=\cos^{-1}(\nu_{\mathrm{BCS}}/\nu_0)$ and $\nu_{\mathrm{BCS}}/\nu_0$ is the BCS density of states on the island.  This analysis predicts a reasonable $s$-wave like proximity induced gap $E_g$ in the normal region, see Fig.~\ref{fig:Figure4}(a), albeit with a scale set by the Thouless energy, $E_g\sim E_{\mathrm{Th}}$, rather than a superconducting gap $\Delta$. However it contains no specific information about the microscopic surface state structure of the material, which is neglected in the semiclassical disordered limit. For the typical known values of $v_F\simeq 5\times10^5$ m/s and $l\simeq80$ nm which are specific to Bi$_2$Se$_3$ surface states, and $R\simeq500$ nm, one estimates a proximity induced Thouless gap $E_{g}\lesssim1$ meV to be in a proper parameter regime when compared with $\Delta$. Based on this modeling the expected superconducting coherence length for disordered surface states is in the range of $\xi=\sqrt{v_Fl/\Delta}\sim200$ nm. However, in order to fit the actual spatial profile of the decay of the gap function $\Delta(x)$ in the Bi$_2$Se$_3$ found in the experiment [see inset in Fig.~\ref{fig:Figure4}(b)] it is necessary to use a fitting parameter of the Thouless scale which is different from what is estimated above. The oscillatory features can also not be reproduced in this limit as, although in principle the resulting LDoS in Eq.~\eqref{LDoS-Usadel} is oscillating, it follows from Eq.~\eqref{theta} that the oscillation and decay scale of the Bessel function are controlled by the same parameter.

We then considered the purely ballistic limit of the proximity effect relevant for the topologically nontrivial surface states. Due to technical complications with the underlying equations specific to the spherical geometry, we solved a geometrically simpler planar model. Namely, we have a two dimensional plane consisting of surface states of a three dimensional topological insulator with $s$-wave pairing in one half of the plain, $x<0$. At the mean-field level this is described by Gor'kov's equation:
\begin{equation}\label{gti}
\begin{pmatrix}
\im\omega_n-H & \im{\bm \sigma}^y\Delta_x\\
-\im{\bm \sigma}^y\Delta^\dagger_x & -\im\omega_n-H^* 
\end{pmatrix}\begin{pmatrix}
G_{n,k_y}(x,x')\\
F^\dagger_{n,k_y}(x,x')
\end{pmatrix}
=
\begin{pmatrix}
\delta(x-x')\\
0
\end{pmatrix}\,,
\end{equation}
with a Hamiltonian
\begin{eqnarray}
H^{(*)}&=&\frac{v_{Fx}}{2}\begin{pmatrix}
0&\pm\hat{k}_x-\im k_y\\
\pm\hat{k}_x+\im k_y&0
\end{pmatrix}+\textrm{H.c}\,,
\end{eqnarray}
which describes the linearized version of the 2D surface states of a 3D topological insulator \cite{Zhang-BiSe-BiTe}.
$\Delta_x=\Delta\Theta(-x)$ is the $s$-wave pairing in the superconducting region with $\Theta$ the Heaviside theta function as before and $v_{Fx}=v_{FS}\Theta(-x)+v_{FT}\Theta(x)$. $G_{n,k_y}(x,x')$ and $F^\dagger_{n,k_y}(x,x')$ are the normal and anomalous Green's functions respectively in the Matsubara representation with fermionic Matsubara frequencies $\omega_n$. Assuming a perfect interface, the boundary conditions are, $G^T|_{x=0^+}=G^S|_{x=0^-}$ and $F^{\dagger T}|_{x=0^+}=F^{\dagger S}|_{x=0^-}$ and with the proper bulk values of the Green's functions as $x\to\pm\infty$. We have used the superscripts $T$ and $S$ for the topological insulator surface and superconductor regions. The local density of states, integrated over the moment $k_y$ transverse to the boundary, is
\begin{equation}\label{LDoS-Gorkov}
\nu(x,\omega)=-\frac{1}{\pi}\Im \int_{-\infty}^{+\infty}\frac{\ud k_y}{2\pi}\tr\left[G^{T,S}_{n,k_y}(x,x)\right]_{i\omega_n\to\omega+\im\delta}\,,
\end{equation}
with $G^{T,S}$ taken as appropriate depending on the position of $x$ and the $\tr$ performed over spin. By solving the Gor'kov equations one finds for the trace of the Green's function $\tr G^S_{n,k_y}(x,x)=(\omega_n/pv^2_{FS})(1-e^{-2ipx})$ where $(v_{FS}p)^2=-[\Delta^2+\omega^2_n+(v_{FS}k_y)^2]$. The expression for $\tr G^T$ is similar. The resulting local density of states is then found from Eq.~\eqref{LDoS-Gorkov} upon an analytical continuation and integration over the transverse momentum 
\begin{equation}\label{LDoS-Tomasch}
\nu(x,\omega)=\frac{|\omega|\Theta(|\omega|-\Delta_x)}{2\pi v_{Fx}^2}\left[1-J_0\left(\frac{2x\sqrt{\omega^2-\Delta_x^2}}{v_{Fx}}\right)\right]\,.
\end{equation}
Here $J_0$ is the Bessel function of the first kind. The bulk density of states in the normal part is $\nu_0(\omega)=|\omega |/2\pi v_{FT}^2$, and in the superconductor we find the bulk density of states $\nu_{S0}(\omega)=|\omega|\Theta(|\omega|-\Delta)/2\pi v_{FS}^2$. Eq.~\eqref{LDoS-Tomasch} displays Friedel-like oscillations induced in the LDoS in the normal side of the junction. The energy scale of these oscillations, $v_{FT}/2x$, are of the same order of magnitude as those we see experimentally for a typical distance in the range of few hundreds of nanometers. On the superconducting side, Eq.~\eqref{LDoS-Tomasch} implies oscillatory LDoS with Tomasch-like functional dependence on energy and position, which physically originates from quasiparticle scattering as induced by a nonuniform superconducting order parameter.
Even though this modeling was done for a different geometry then that of our system, we believe that these oscillations are generic. Additional analysis shows that for the spherical symmetry Gor'kov equations \eqref{gti} lead to a LDoS of a more complicated functional form then Eq.~\eqref{LDoS-Tomasch} with Bessel functions of different harmonics but the essence of the effect remains the same.  
  
\textbf{\textit{Summary}}. We used superconducting islands deposited on the surface of topological insulator to induce local superconducting correlations into the surface states by virtue of the proximity effect. 
Superconducting gaps are revealed in the local density of states measurements by cryogenic scanning tunneling microscopy. The gap function is studied as a function of distance away from the island and along the surface state. From these spatially resolved measurements we were able to estimate the range and scale of the proximity effect. We also observed additional oscillatory features in the STM spectra and provided evidence for the signatures of the Dirac cone from the local density of states measurements on superconducting islands. Our mesoscopic system will pave the way to further study direct superconducting and inverse topological proximity effects.  

\textbf{\textit{Acknowledgements}}.  We thank Matthias Muenks for key contributions in developing the sample preparation methods. We thank Christopher Reeg for discussions concerning Ref.~\cite{Reeg}. Support for this research at Michigan State University (N.S.) was provided by the Institute for Mathematical and Theoretical Physics with funding from the office of the Vice President for Research and Graduate Studies. This work at University of Wisconsin-Madison (A.L.) was financially supported in part by NSF Grants No. DMR-1606517, ECCS-1560732, and BSF Grant No. 2014107, and the Wisconsin Alumni Research Foundation. The work at Northwestern University was supported by the Department of Energy, Office of Science Basic Energy Sciences grant DE-SC0014520. I.M.D. acknowledges fellowship support from the Department of Education through GAANN award USDE P200A140215 and from the Peter Schroeder Endowment.

\appendix

\section{Extracting the value of $\Delta$ from the experimental data}

In this experiment we measured the proximity-induced superconducting gap parameter $\Delta$ on $\textrm{Bi}_2 \textrm{Se}_3$, and on proximity coupled superconducting islands of PbBi. We measured the transition temperature of a 400 nm thick film of $\textrm{Pb}_{0.3}\textrm{Bi}_{0.7}$ using a standard four-probe transport measurement; we found the value of the transition temperature to be 8.3 K, which compares well to the nominal value of 8.2 K [See Ref. 38].  $\Delta$ is essentially the energy difference of the decoherence peaks in the DOS spectra divided by two with a nominal value of 3.65 meV [See Ref. 38]. However, energy broadening effects, both thermal and mean-free-path, shift the locations of the peaks; moreover, as we found a coexistence of the Dirac cone with the induced superconducting gaps, care must be taken to account for any additional distortions in the location of the peaks.

\begin{figure}[htpb!]
\centering
\includegraphics*[width=0.95\linewidth]{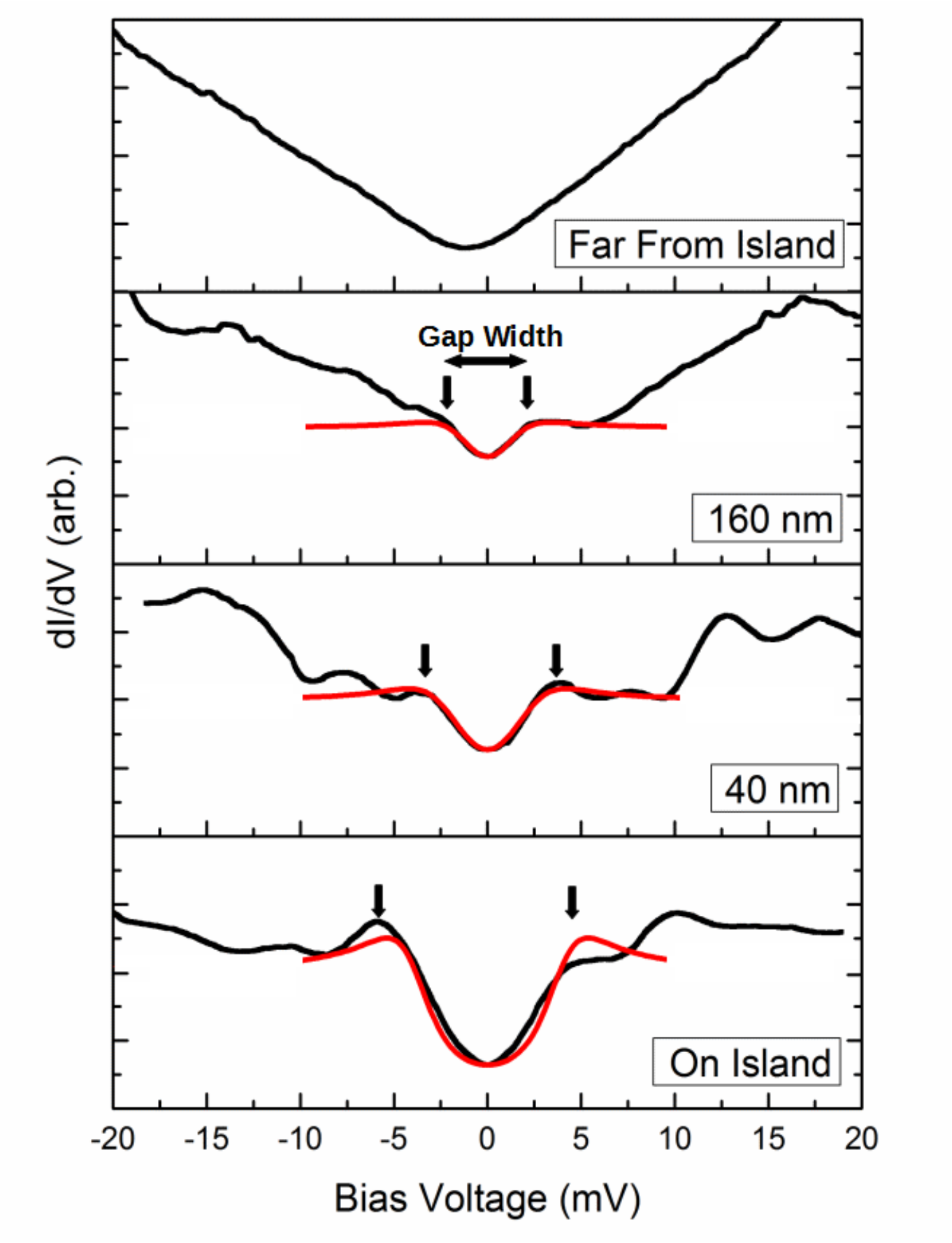}
\caption{(Color online). Experimental data of the superconducting proximity effect measurement of superconductor PbBi in contact with topological insulator $\textrm{Bi}_2 \textrm{Se}_3$ fitted with calculated s-wave superconducting DOS curves.  The arrows denote the locations of $\Delta$ used when calculating the gap width after linear subtraction was used to find the decoherence peaks.}  
\label{fig:Figure1_Supp}
\end{figure}

To account for energy broadening effects, we fit several measured curves with s-wave BCS superconducting DOS curves which were appropriately broadened. This allowed us to account for the small energy difference between the measured peak locations and the underlying value of $\Delta$. With regard to distortions caused by the approximately linear background, we accounted for this by subtracting a linear background for the positive and negative branches of the curves at low energies and then identifying the voltage of the respective decoherence peaks. Of course, noise is present which limits the precision of this procedure. We estimate that the values of $\Delta$ are accurate to about $\pm10\%$. Fig. \ref{fig:Figure1_Supp} shows three measured curves (with no background subtractions) superposed with the calculated s-wave fits.

\section{Calculations for the dirty limit}\label{sec_usadel}

We consider the Usadel equation for a two dimensional system at an energy $\omega$:
\begin{equation}
D\nabla^2\theta(\omega,\ver)+2\im\omega\sin\theta(\omega,\ver)+\Delta(\ver)\cos\theta(\omega,\ver)=0\,.
\end{equation}
$D=v_{FS}\ell$ is the diffusion constant, where $v_{FS}$ is the Fermi velocity in the superconductor and $\ell$ is the mean free path. $\Delta(\ver)$ is the strength of the s-wave pairing as a function of coordinate $\ver$.
The local density of states is
\begin{equation}
\frac{\nu(\omega,\ver)}{\nu_0}=\Re\cos\theta(\omega,\ver)\,,
\end{equation}
rescaled to the density of states in the absence of the pairing $\nu_0$. The pairing term is found from 
\begin{equation}
\F(\omega,\ver)=\sin\theta(\omega,\ver)\,,
\end{equation}
the anomalous Green's function.

In Fig.~(4) in the main paper we plot the gap function extracted from the analytical result to the Usadel equation, compared to the experimental results. The normalized gap function is found from
\begin{equation}
\Delta(x)=\int_{-\infty}^\infty\ud\omega\Re\sin\theta(\omega,x)\,.
\end{equation}

We are interested in modeling a superconducting island surrounded by a normal material and hence we will work in polar coordinates $\ver=(r,\theta)$. Assuming circular symmetry of the system and with $\Delta(\ver)=\Delta\Theta(R-r)$, where $\Theta$ is the Heaviside theta function, i.e.~ we have a circular superconductor of radius $R$ surrounded by a normal metal. Let us rescale position as $x=r/R$. We then have, in the normal metal,
\begin{equation}
\frac{\ud^2\theta}{\ud x^2}+\frac{1}{x}\frac{\ud\theta}{\ud x}+\frac{\im\omega}{E_{\rm Th}}\sin\theta=0\,,
\end{equation}
where $E_{\rm Th}=D/R^2$ is the Thouless energy. $\theta_0$ is determined from the boundary condition at $x=1$ and is $\theta_0=\cos^{-1}(\nu_{\rm BCS}/\nu_0)$ with $\nu_{\rm BCS}/\nu_0$ the BCS density of states on the island:
\begin{equation}\label{bcss}
\frac{\nu_{\rm BCS}(\omega)}{\nu_0}=\Re\frac{|\omega|}{\sqrt{\omega^2-\Delta^2}}\,.
\end{equation}

\section{Calculations in the clean limit}

We give here more details for the calculation of the density of states for the surface states of a topological insulator in the clean limit. In this case we have a two dimensional plain consisting of surface states of a three dimensional topological insulator with s-wave pairing in one half of the plane, $x<0$. At the mean-field level this is described by Gorkov's equation:
\begin{equation}\label{gtiapp}
\begin{pmatrix}
\im\omega_n-H & \im{\bm \sigma}^y\Delta_x\\
-\im{\bm \sigma}^y\Delta^\dagger_x & -\im\omega_n-H^* 
\end{pmatrix}\begin{pmatrix}
G_{n,k}(x,x')\\
\F^\dagger_{n,k}(x,x')
\end{pmatrix}
=
\begin{pmatrix}
\delta(x-x')\\
0
\end{pmatrix}\,,
\end{equation}
with a Hamiltonian
\begin{eqnarray}
H^{(*)}&=&\frac{v_{Fx}}{2}\begin{pmatrix}
0&\pm\hat{k}_x-\im k\\
\pm\hat{k}_x+\im k&0
\end{pmatrix}+\textrm{H.c}\,,
\end{eqnarray}
which describes the linearized form for the 2D surface states of a 3D topological insulator with $k$ the momentum parallel to the boundary.
$\Delta_x=\Delta\Theta(-x)$ is the s-wave pairing in the superconducting region with $\Theta$ the Heaviside theta function as before and $v_{Fx}=v_{FS}\Theta(-x)+v_{FT}\Theta(x)$. $G_{n,k}(x,x')$ and $\F^\dagger_{n,k}(x,x')$ are the normal and anomalous Green's functions respectively in the Matsubara representation with fermionic Matsubara frequencies $\omega_n$:
\begin{eqnarray}\label{matg}
G(x,\tau;x',\tau')&=&-\langle\T_\tau\psi_{\sigma}(x,\tau)\psi^\dagger_{\sigma'}(x',\tau')\rangle\\\nonumber
&=&T\sum_n\e^{-\im \omega_n(\tau-\tau')}G_n(x,x')\,,
\end{eqnarray}
and
\begin{eqnarray}\label{matf}
\F^\dagger(x,\tau;x',\tau')&=&-\langle\T_\tau\psi^\dagger_{\sigma}(x,\tau)\psi^\dagger_{\sigma'}(x',\tau')\rangle\\\nonumber
&=&T\sum_n\e^{-\im \omega_n(\tau-\tau')}\F^\dagger_n(x,x')\,,
\end{eqnarray}
with $\T_\tau$ time ordering along the imaginary time axis and $\psi_{\sigma}(x,\tau)$ a Heisenberg operator.

The boundary conditions are
\begin{eqnarray}
G^T|_{x=0^+}&=&G^S|_{x=0^-}\,,\textrm{ and}\nonumber\\
\F^{\dagger T}|_{x=0^+}&=&\F^{\dagger S}|_{x=0^-}\,.
\end{eqnarray}
We have used the superscripts $T$ and $S$ for the Rashba and superconductor regions, and the Green's functions must vanish at $\pm\infty$. For the local density of states we want 
$G^{S,T}_{n,k}(x,x)$.

First let us note the momenta. In the superconducting region we have
\begin{equation}\label{eqp}
v_{FS}^2(p^n_k)^2=-(\Delta^2+\omega_n^2+v_{FS}^2k^2)\,,
\end{equation}
with $\Im p>0$, which is imposed by the boundary condition at $x\to-\infty$.
In the topological insulator surface states we have
\begin{equation}\label{eqq}
v_{FT}q^n_k=\im\sqrt{\omega_n^2+v_{FT}^2k^2}\,,
\end{equation}
taking the positive square root, which is imposed by the boundary condition at $x\to\infty$.
Finally let us define
\begin{equation}\label{eqg}
\gamma^n_k=\frac{v_{FT}}{\omega_n}(k-\im q^n_k)\,,
\end{equation}
and
\begin{equation}\label{eqgb}
\beta^n_k=\frac{v_{FS}}{\omega_n}(k-\im p^n_k)\,.
\end{equation}

The bulk solution, for $x,x'>0$, is
\begin{eqnarray}
G^{T0}_{n,k}(x-x')&=&-\frac{\im \e^{\im q^n_k|x-x'|}}{2 v_{FT}^2}\times
\\&&\nonumber
\left[\frac{\im\omega_n-v_{FT}k{\bm\sigma}^y}{q^n_k}-v_{FT}\sgn(x-x'){\bm\sigma}^x\right]\,.
\end{eqnarray}
We find
\begin{equation}
G^{T}_{n,k}(x,x')=G^{T0}_{n,k}(x-x')+\frac{\omega_n\e^{\im q^n_k (x+x')}}{2v_{FT}^2q^n_k}
\begin{pmatrix}
-1&\gamma^n_k\\
\frac{1}{(\gamma^n_k)^*}&-1
\end{pmatrix}\,.
\end{equation}
In particular we then find
\begin{equation}\label{gtxx}
G^{T}_{n,k}(x,x)=\frac{\omega_n+\im v_{FT}k{\bm\sigma}^y}{2 v_{FT}^2q^n_k}+\frac{\omega_n\e^{2\im q^n_k x}}{2v_{FT}^2q^n_k}\begin{pmatrix}
-1&\gamma^n_k\\
\frac{1}{(\gamma^n_k)^*}&-1
\end{pmatrix}\,.
\end{equation}

The bulk solution, for $x,x'<0$, is
\begin{eqnarray}
G^{S0}_{n,k}(x-x')&=&\frac{\im \e^{\im p^n_k|x-x'|}}{2 v_{FS}^2}\times
\\&&\nonumber
\left[\frac{v_{FS}k{\bm\sigma}^y-\im\omega_n}{p}+v_{FS}\sgn(x-x'){\bm\sigma}^x\right]\,,
\end{eqnarray}
and we find
\begin{equation}
G^{S}_{n,k}(x,x')=G^{S0}_{n,k}(x-x')+\frac{\omega_n \e^{-\im p^n_k(x+x')}}{2v_{FS}^2 p^n_k}\begin{pmatrix}
-1&\beta^n_k\\
\frac{1}{(\beta^n_k)^*}&-1
\end{pmatrix}\,.
\end{equation}
In particular we will want
\begin{equation}\label{gsxx}
G^{S}_{n,k}(x,x)=
\frac{\omega_n+\im v_{FS}k{\bm\sigma}^y}{2 v_{FS}^2p^n_k}+\frac{\omega_n\e^{-2\im p^n_kx}}{2v_{FS}^2 p^n_k}\begin{pmatrix}
-1&\beta^n_k\\
\frac{1}{(\beta^n_k)^*}&-1
\end{pmatrix}\,.
\end{equation}

\subsection{Local density of states}

The local density of states is defined by
\begin{equation}
\nu(x,\omega)=-\frac{1}{\pi}\Im \int_{-\infty}^\infty\frac{\ud k}{2\pi}\tr\left.G^{T,S}_{n,k}(x,x)\right|_{i\omega_n\to\omega+\im\delta}\,,
\end{equation}
with $G^{T,S}$ taken as appropriate depending on the position of $x$ and $\tr$ being the trace over the spin.
\begin{figure}
\includegraphics*[width=0.95\columnwidth]{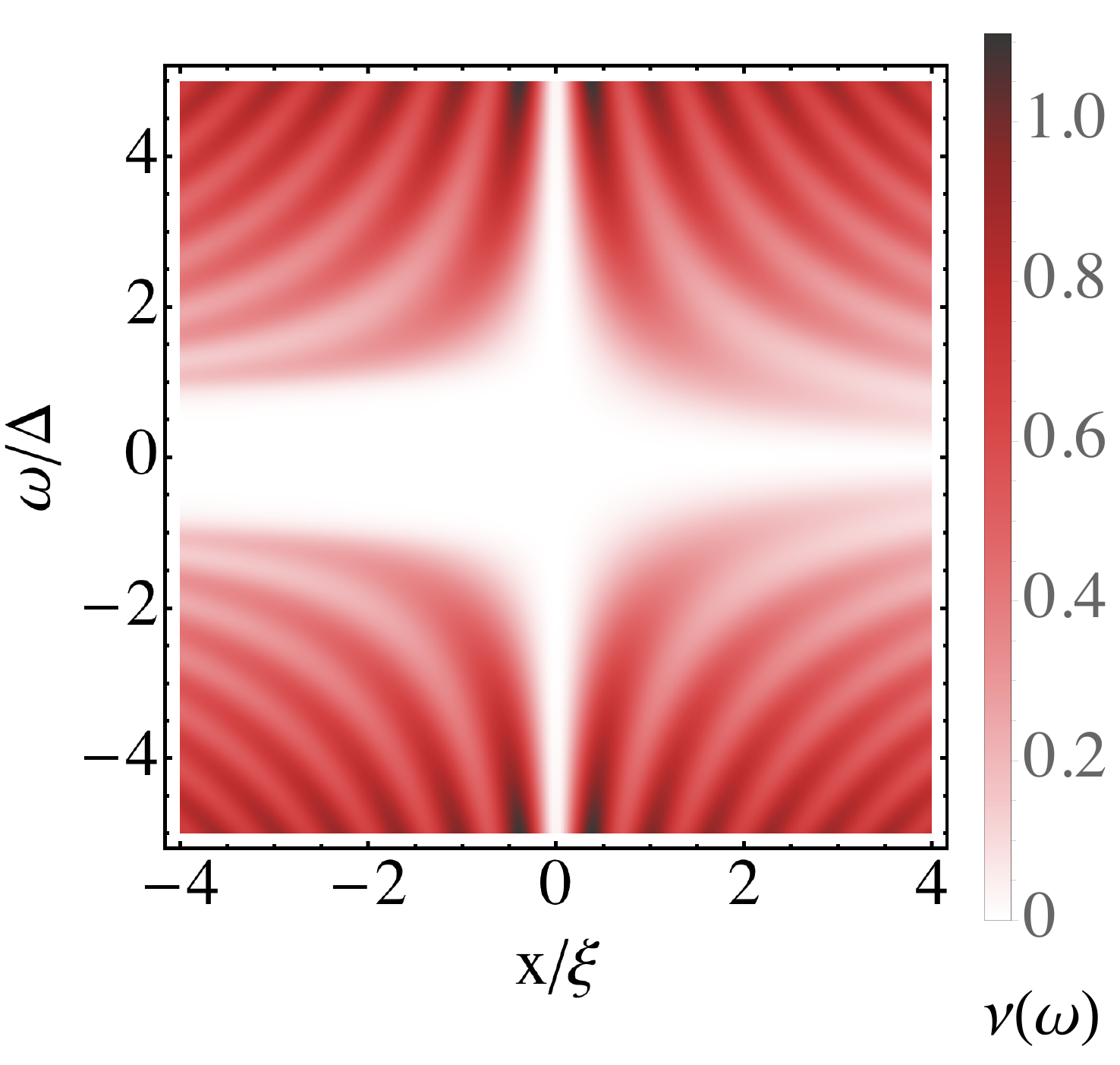}
\caption{The local density of states, in arbitrary units, as a function of energy for the topological insulator system with $\Delta=1$ and $v_{FT}=v_{FS}=1$. A phenomenological damping of magnitude $\Gamma=0.25\Delta$ is included. The position is measured in units of the superconducting coherence length $\xi=v_{FS}/\Delta$.}
\label{ldosti2d}
\end{figure}
\begin{figure}
\includegraphics*[width=0.95\columnwidth]{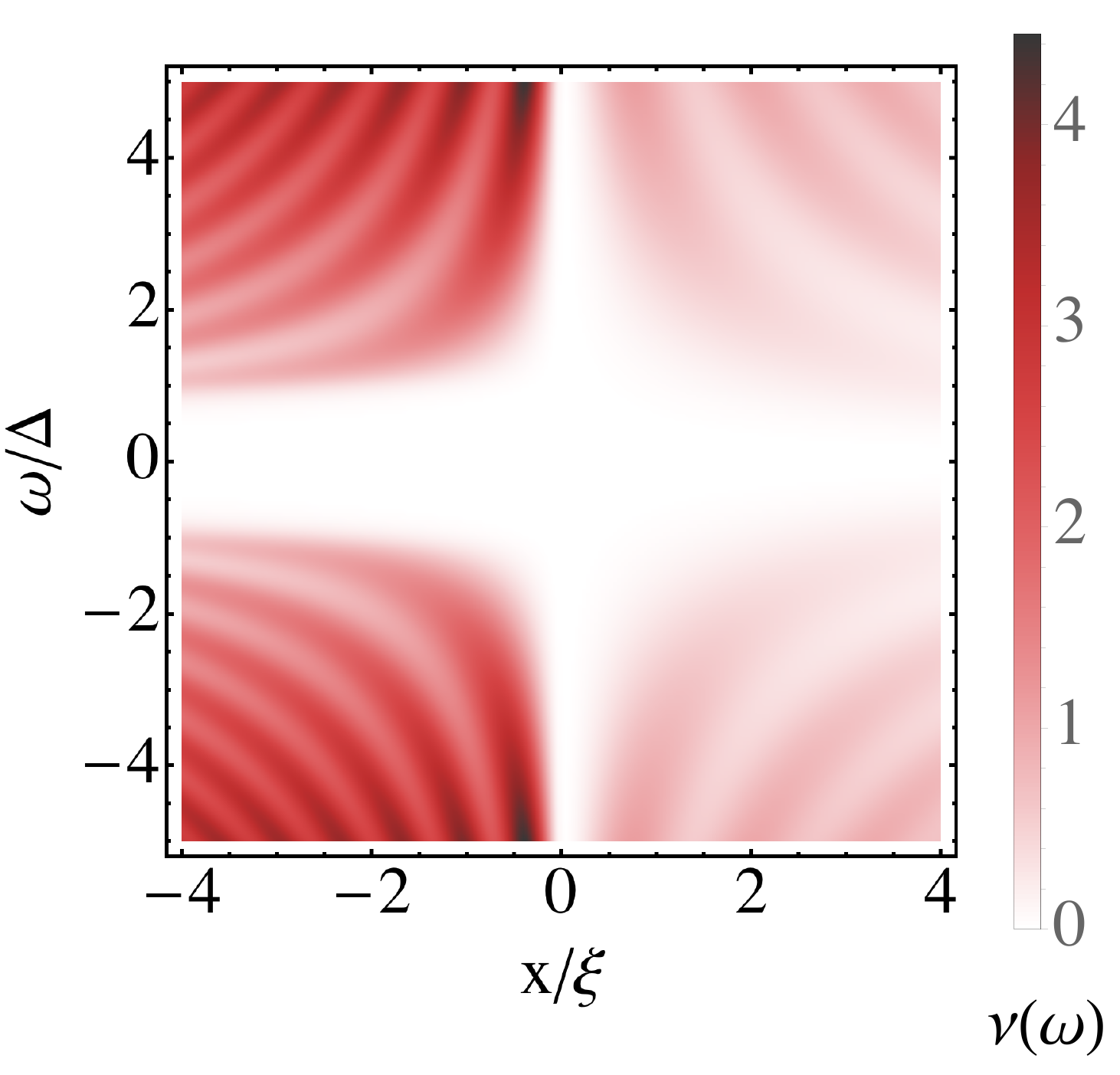}
\caption{The local density of states, in arbitrary units, as a function of energy for the topological insulator system with $\Delta=1$ and $v_{FT}=2v_{FS}=1$. A phenomenological damping of magnitude $\Gamma=0.25\Delta$ is included. The position is measured in units of the superconducting coherence length $\xi=v_{FS}/\Delta$.}
\label{ldosti2d2}
\end{figure}

The bulk density of states is
\begin{equation}\label{diracdos}
\nu_0(\omega)=\frac{|\omega |}{2\pi v_{Fx}^2}\Theta(|\omega|-\Delta_x)\,.
\end{equation}
From Eqn.~\eqref{gtxx} we find
\begin{equation}
\tr G^{R}_{n,k}(x,x)=\frac{\omega_n}{v_{FT}^2q}\left(1-\e^{2\im q x}\right)\,.
\end{equation}
Substituting this into the definition for the local density of states, we find
\begin{eqnarray}
\frac{\nu(x>0,\omega)}{\nu_{0}(\omega)}&=&1-
\int_{-1}^{1}\frac{\ud y}{\pi}\frac{\cos\left[\frac{2x|\omega|}{v_{FT}}\sqrt{1-y^2}\right]}{\sqrt{1-y^2}}\nonumber\\
&=&1-J_0\left(\frac{2x|\omega|}{v_{FT}}\right)\,.
\end{eqnarray}
We have used a substitution in the integral of $y=v_{FT}k/\omega$. The limits of the integral are determined by when the imaginary part of the Green's function is non-zero and 
$J_0$ is the Bessel function of the first kind.

In the superconducting region 
we find from Eqn.~\eqref{gsxx} that
\begin{equation}
\tr G^{S}_{n,k}(x,x)=\frac{\omega_n}{v_{FS}^2p}\left(1-\e^{-2\im p x}\right)\,.
\end{equation}
Substituting this into the definition for the local density of states:
\begin{eqnarray}
\frac{\nu(x< 0,\omega)}{\nu_{S0}(\omega)}&=&1-
\int_{-1}^{1}\frac{\ud y}{\pi}\frac{\cos\left[\frac{2x\sqrt{\omega^2-\Delta^2}}{v_{FS}}\sqrt{1-y^2}\right]}{\sqrt{1-y^2}}\nonumber\\
&=&1-J_0\left(\frac{2x|\sqrt{\omega^2-\Delta^2}|}{v_{FS}}\right)\,.
\end{eqnarray}
We have used a substitution in the integral of $y=v_{FS}k/\Omega$. Again, the limits of the integral are determined by when the imaginary part of the Green's function is non-zero.
At the boundary the local density of states satisfies
\begin{equation}
\nu(x=0^+,\omega)=\nu(x=0^-,\omega)=0\,.
\end{equation}
Some examples are given in Figs.~\ref{ldosti2d} and \ref{ldosti2d2} for systems with a change in the Fermi velocity across the boundary and for a uniform Fermi velocity across the whole system.


\begin{thebibliography}{99}

\bibitem{Hasan-Kane}
M. Z. Hasan and C. L. Kane, Rev. Mod. Phys. \textbf{82}, 3045 (2010). 

\bibitem{Hsieh-BiSb}
D. Hsieh, D. Qian, L. Wray, Y. Xia, Y. S. Hor, R. J. Cava, and M. Z. Hasan, 2008, Nature (London) \textbf{452}, 970.

\bibitem{Xia-BiSe}
Y. Xia, D. Qian, D. Hsieh, L. Wray, A. Pal, H. Lin, A. Bansil, D. Grauer, Y. S. Hor, R. J. Cava, and M. Z. Hasan, Nat. Phys. \textbf{5}, 398 (2009).

\bibitem{Zhang-BiSe-BiTe}
H. Zhang, C.-X. Liu, X.-L. Qi, X. Dai, Z. Fang, and S.-C. Zhang, Nat. Phys. \textbf{5}, 438 (2009).

\bibitem{Fu-Kane}
L. Fu and C. L. Kane, Phys. Rev. Lett. \textbf{100}, 096407 (2008).  

\bibitem{Linder}
J. Linder, Y. Tanaka, T. Yokoyama, A. Sudbo, and N. Nagaosa,
Phys. Rev. Lett. \textbf{104}, 067001 (2010).

\bibitem{Potter}
A. C. Potter and P. A. Lee, Phys. Rev. B \textbf{83}, 184520 (2011). 

\bibitem{Lababidi}
M. Lababidi and E. Zhao, Phys. Rev. B \textbf{83}, 184511 (2011).

\bibitem{Ioselevich}
P. A. Ioselevich, P. M. Ostrovsky, and M. V. Feigel'man, Phys. Rev. B \textbf{86}, 035441 (2012).  

\bibitem{Tkachov}
G. Tkachov, Phys. Rev. B \textbf{87}, 245422 (2013). 

\bibitem{Reeg}
C. R. Reeg and D. L. Maslov, Phys. Rev. B \textbf{92}, 134512 (2015). 

\bibitem{Sasaki}
S. Sasaki, M. Kriener, K. Segawa, K. Yada, Y. Tanaka, M. Sato, and Y. Ando, Phys. Rev. Lett. \textbf{107}, 217001 (2011). 

\bibitem{Wang-BiSe-NbSe-STM}
M.-X. Wang, C. Liu, J.-P. Xu, F. Yang, L. Miao, M.-Y. Yao, C. L. Gao, C. Shen, X. Ma, X. Chen, Z.-A. Xu, Y. Liu, S.-C. Zhang, D. Qian, J.-F. Jia, Q.-K. Xue, Science \textbf{336}, 52 (2012).

\bibitem{Wang-BiSe-BiSrCaCuO-ARPES}
E. Wang, H. Ding, A.V. Fedorov, W. Yao, Z. Li, Y.-F. Lv, K. Zhao, L.-G. Zhang, Z. Xu, J. Schneeloch, R. Zhong, S.-H. Ji,  L. Wang,	K. He, X. Ma, G. Gu,	H. Yao, Q.-K. Xue,  X. Chen, and S. Zhou, Nat. Phys. \textbf{9}, 621 (2013). 

\bibitem{Koren-BiSe-NbN}
G. Koren, T. Kirzhner, Y. Kalcheim, and O. Millo, Europhys. Lett. \textbf{103}, 67010 (2013).  

\bibitem{Xu-BiTe-STM}
J.-P. Xu, C. Liu, M.-X. Wang, J. Ge, Z.-L. Liu, X. Yang, Y. Chen, Y. Liu, Z.-A. Xu, C.-L. Gao, D. Qian, F.-C. Zhang, and J.-F. Jia, Phys. Rev. Lett. \textbf{112}, 217001 (2014).

\bibitem{Sacepe}
B. Sacepe, J. B. Oostinga, J. Li, A. Ubaldini, N. J. G. Couto, E. Giannini, and A. F. Morpurgo, Nat. Commun. \textbf{2}, 575 (2011).

\bibitem{Veldhorst}
M. Veldhorst, M. Snelder, M. Hoek, T. Gang, V. K. Guduru, X. L. Wang, U. Zeitler, W. G. Van der wiel, A. A. Golubov, H. Hilgenkamp, and A. Bririkman, Nat. Mater. \textbf{11}, 417 (2012).

\bibitem{Qu}
F. M. Qu, F. Yang, J. Shen, Y. Ding, J. Chen, Z. Q. Ji, G. G. Liu, J. Fan, X. N. Jing, C. L. Yang, and L. Lu, Sci. Rep. \textbf{2}, 339 (2012).

\bibitem{Bestwick}
J. R. Williams, A. J. Bestwick, P. Gallagher, Seung Sae Hong, Y. Cui, Andrew S. Bleich, J. G. Analytis, I. R. Fisher, and D. Goldhaber-Gordon, Phys. Rev. Lett. \textbf{109}, 056803 (2012). 

\bibitem{Mason}
S. Cho, B. Dellabetta, A. Yang, J. Schneeloch, Z. Xu, T. Valla, G. Gu,  M. J. Gilbert, and N. Mason, Nat. Commun. \textbf{4}, 1689 (2013). 

\bibitem{Kurter-PRB}
C. Kurter, A. D. K. Finck, P. Ghaemi, Y. S. Hor, and D. J. Van Harlingen, Phys. Rev. B \textbf{90}, 014501 (2014).

\bibitem{Kurter-NC}
C. Kurter, A. D. K. Finck, Y. S. Hor, D. J. Van Harlingen, Nat. Commun. \textbf{6}, 7130 (2015).

\bibitem{Sochnikov}
I. Sochnikov, L. Maier, C. A. Watson, J. R. Kirtley, C. Gould, G. Tkachov, E. M. Hankiewicz, C. Br\"une, H. Buhmann, L. W. Molenkamp, and K. A. Moler, Phys. Rev. Lett. \textbf{114}, 066801 (2015). 

\bibitem{Stehno}
M. P. Stehno, V. Orlyanchik, C. D. Nugroho, P. Ghaemi, M. Brahlek, N. Koirala, S. Oh, and D. J. Van Harlingen, Phys. Rev. B \textbf{93}, 035307 (2016)

\bibitem{Zhang-PRB11}
D. Zhang, J. Wang, A. M. DaSilva, J. S. Lee, H. R. Gutierrez, M. H. W. Chan, J. Jain, and N. Samarth, 
Phys. Rev. B \textbf{84}, 165120 (2011). 

\bibitem{Finck-PRX}
A.D.K. Finck, C. Kurter, Y.S. Hor, D.J. Van Harlingen, Phys. Rev. X \textbf{4}, 041022 (2014).

\bibitem{Finck-arXiv}
A.D.K. Finck, C. Kurter, E.D. Huemiller, Y.S. Hor, D.J. Van Harlingen, preprint arXiv:1503.06898. 

\bibitem{Zareapour}
P. Zareapour, A. Hayat, S. Y. Zhao, M. Kreshchuk, A. Jain, D. C. Kwok, N. Lee, S. W. Cheong, Z. J. Xu, A. Yang, G. D. Gu, S. Jia, R. J. Cave, and K. S. Burch, Nat. Commun. \textbf{3}, 1056 (2012).

\bibitem{Wray}
L. A. Wray, S.-Y. Xu, Y. Xia, Y. S. Hor, D. Qian, A. V. Fedorov, H. Lin, A. Bansil, R. J. Cava, and M. Zahid Hasan, Nat. Phys. \textbf{6}, 855 (2010). 

\bibitem{Hor}
Y. S. Hor, A. J. Williams, J. G. Checkelsky, P. Roushan, J. Seo, Q. Xu, H. W. Zandbergen, A. Yazdani, N. P. Ong, and R. J. Cava, Phys. Rev. Lett. \textbf{104}, 057001 (2010). 

\bibitem{Tikhonov}
M. V. Feigel'man, M. A. Skvortsov, K. S. Tikhonov,  JETP Lett. \textbf{88}, 862 (2008). 

\bibitem{Mason-SNS}
S. Eley, S. Gopalakrishnan, P. M. Goldbart, N. Mason, Nat. Phys. \textbf{8}, 59 (2012).

\bibitem{degennes} P. G. Gennes, in Superconductivity Of Metals And Alloys, (Westview Press, 1999).

\bibitem{DDG} G. Deutscher and P. G. Gennes, in Superconductivity, edited by R. D. Parks (Dekker, NY, 1969).

\bibitem{Wolfram}
T. Wolfram, Phys. Rev. \textbf{170}, 481 (1968). 

\bibitem{Megan} 
M. Romanowich, M.-S. Lee, D.-Y. Chung, S. D. Mahanti, M. G. Kanatzidis, and S. H. Tessmer, Phys. Rev. B \textbf{87}, 085310 (2013).

\bibitem{Adler} J. G. Adler and S. C. Ng, Can. J. of Phys. \textbf{43}, 594 (1965).

\bibitem{Bianchi}
M. Bianchi, D. Guan,	S. Bao, J. Mi, B. Brummerstedt Iversen, P. D.C. King, and P. Hofmann
Nat. Commun. \textbf{1}, 128 (2010).

\bibitem{SN-STM}
N. Moussy, H. Courtois and B. Pannetier, Europhys. Lett., \textbf{55} (6), 861 (2001). 

\bibitem{AL}
A. Levchenko, Phys. Rev. B \textbf{77}, 180503(R) (2008). 	
    
\end{thebibliography}
\end{document}